\documentstyle[12pt]{article}
\input epsf

\input epsf

\begin{document}

\renewcommand{\thesection}{\arabic{section}.} 
\renewcommand{\theequation}{\thesection \arabic{equation}}
\newcommand{\scs}{\setcounter{equation}{0} \setcounter{section}}
\def\req#1{(\ref{#1})}
\newcommand{\be}{\begin{equation}} \newcommand{\ee}{\end{equation}} 
\newcommand{\ba}{\begin{eqnarray}} \newcommand{\ea}{\end{eqnarray}} 
\newcommand{\la}{\label} \newcommand{\nb}{\normalsize\bf} 
\newcommand{\lb}{\large\bf} \newcommand{\vol}{\hbox{Vol}}
\newcommand{\bb} {\bibitem} \newcommand{\np} {{\it Nucl. Phys. }} 
\newcommand{\pl} {{\it Phys. Lett. }} 
\newcommand{\pr} {{\it Phys. Rev. }} \newcommand{\mpl} {{\it Mod. Phys. Lett. }}
\newcommand{\sg}{{\sqrt g}} \newcommand{\sqhat}{{\sqrt{\hat g}}}
\newcommand{\sqphi}{{\sqrt{\hat g}} e^\phi} 
\newcommand{\sqalpha}{{\sqrt{\hat g}}e^{\alpha\phi}}
\newcommand{\tp}{\cos px\ e^{(p-{\sqrt2})\phi}} \newcommand{\stwo}{{\sqrt2}}
\newcommand{\tr}{\hbox{tr}}

\begin{titlepage}
\renewcommand{\thefootnote}{\fnsymbol{footnote}}

\hfill CERN-TH/2001-087

\hfill hep-th/0104131

\vspace{.4truein}
\begin{center}
 {\LARGE Brane Supersymmetry Breaking and the 

\vskip4mm

 Cosmological Constant: Open Problems
}
 \end{center}
\vspace{.7truein}

 \begin{center}

Christof Schmidhuber
\footnote{christof.schmidhuber@cern.ch}

 \vskip5mm

 {\it CERN, Theory Division, 1211 Gen\`eve 23}

 \end{center}

\vspace{.5truein}
\begin{abstract}

\noindent
It has recently been argued that non--BPS brane world scenarios
can reproduce the small value of the cosmological constant
that seems to have been measured.
Objections against this proposal are discussed and necessary (but not
sufficient) conditions are stated under which it may work. 
At least $n=2$ extra dimensions are needed. Also, the mass matrix 
in the supergravity sector must satisfy $Str\ {\cal M}^2=0$.
Moreover, the proposal can be ruled out experimentally if Newton's 
constant remains unchanged down to scales of $10$ $\mu m$.
If, on the other hand, such a ``running Newton constant'' is observed,
it could provide crucial experimental input for
superstring phenomenology.

\end{abstract}
 \renewcommand{\thefootnote}{\arabic{footnote}}
 \setcounter{footnote}{0}
\end{titlepage}

\subsection*{1. Introduction}\scs{1}

The energy density $\rho$ of the vacuum enters Einstein's
equations in the form of an effective cosmological constant
$\lambda$:
\ba R_{\mu\nu}\ -\ {1\over2}\ g_{\mu\nu}\ R\ =\ \lambda\ g_{\mu\nu}\
\ \ \ \hbox{with}\ \ \ \ \lambda\ =\ 
8\pi G\ \rho\ ,\la{anna}\ea
where $G$ is Newton's constant. Such a cosmological constant $\lambda$
curves four--dimensional space--time, giving it a curvature
radius $R_{curv}$ of order $\lambda^{-{1\over2}}$. 
Only part
 of the curvature of the universe is due to the cosmological constant;
other contributions come from visible and dark matter.
This is accounted for by a factor
$3\Omega_\Lambda$:
$$3\Omega_\Lambda\ R_{curv}^{-2}\ \sim\ \lambda\ .$$
$\Omega_\Lambda$ seems to have been 
measured to be $\Omega_\Lambda\sim{2\over3}$ \cite{bah}.
For a spacially flat universe, which is the case that
seems to be realized in nature, this curvature radius implies an
expanding universe with metric
\ba ds^2\ \sim\ -\ dt^2\ +\ e^{Ht}\ d{\vec x}^2\ \ \ \ \hbox{with}\ \ \ 
3\Omega_\Lambda\ H^2\ \sim\ \lambda .\la{hugo}\ea
The Hubble constant $H$ in our universe is of order
$$H\ \sim\ 10^{-33}eV\ \sim \ (10^{26}m)^{-1}\ .$$
The cosmological constant problem is the question why this
is so small, given that we expect  much larger contributions
to the energy density of the vacuum from Standard Model physics.
Let us briefly recall why we expect such contributions \cite{wei}.

First of all, there are classical effects.
E.g., in the course of electroweak symmetry breaking the Higgs field
is supposed to roll down its potential, thereby changing the
vacuum energy density by an amount of order $(TeV)^4$.
This would give a curvature radius of the universe in the millimeter
range, which is not what we observe. Of course, the Higgs potential
could be shifted so that its Minimum is at zero. Equivalently,
 a bare cosmological
constant could be added in (\ref{anna}) that exactly cancels
the cosmological constant induced by $\rho$. But this is just the fine--tuning
problem: why should the bare cosmological constant be fine--tuned to
exactly cancel the contributions from later phase transitions?

Even if we do find a reason why the minimum of the Higgs potential
should be zero, the problem does not go away. There are other condensates
in the Standard Model, such as the chiral and gluon condensates in
QCD after chiral symmetry breaking. They should contribute an amount
of order $(\Lambda_{QCD})^4$ to the cosmological constant, which would
translate into a curvature radius of the universe in the kilometer range.

And even if one did not believe in chiral and gluon condensates,
there would still remain what is perhaps the most mysterious aspect
of the problem: the cosmological constant does not seem to
receive contributions from the zero--point energy of the Standard Model
fields. A harmonic  oscillator has a ground state energy ${1\over2}\hbar
\omega$. In a field theory we have one oscillator for each 
momentum $k$ with frequency
$$\omega\ =\ \sqrt{k^2+m^2}\ .$$
In the case of the photon (with $m=0$), integrating 
over $k$ gives, for each of the two polarizations, a divergent  contribution
\ba\lambda\ \ =\ 8\pi G\ \int_{|k|\le\Lambda}{d^3k\over(2\pi)^3}{1\over2}
\hbar\omega\ \sim\ {1\over2\pi} {\Lambda^4\over m_{Pl}^2}\la{beate}\ea
to the cosmological constant, where $\hbar G\ =\ m_{Pl}^{-2}$ and
$\Lambda$ is a large--momentum cutoff. How large does
this cutoff $\Lambda$ have to be?
(\ref{beate},\ref{hugo}) tell us that $\Lambda$ is, on a logarithmic scale, half--way
between the Planck mass and the Hubble constant. So $\Lambda$ is several
milli-eV, corresponding to a Compton wave length in the micrometer range.
This is even larger than the wave--length of visible light: we can
see with our bare eyes that there is no such cutoff in nature!

There is one idea that can explain a zero cosmological constant:
supersymmetry. In supersymmetric theories, the contributions from
bosons and fermions to $\lambda$ have opposite signs and
cancel. But in order to explain the smallness of the Hubble constant,
supersymmetry would have to remain unbroken at least up to
scales of the order of the above cutoff $\Lambda$, i.e. up to milli-eV
scales (see figure 1), which is of course ruled out for the Standard Model.

It has been emphasized that the nearly vanishing cosmological
constant should be the main clue 
as to {\it how} nature breaks supersymmetry \cite{wit3}.
Based on \cite{rub, svv, ver2},
it has recently been argued \cite{schm} that supersymmetry breaking
may generate only a tiny cosmological constant, 
 if we assume that we live on a non--BPS 3--brane soliton that is embedded
in a $(4+n)$--dimensional supergravity background. 
This situation can be obtained by compactifying some string theory
on a $(6-n)$--dimensional manifold.
In such ``brane world models,'' there is
 a fundamental difference between
the Standard Model fields, which originate from and live only on
the brane, and the supergravity fields, which
are allowed to propagate everywhere in the bulk.

What has been argued is the following (see \cite{schm}):
\begin{enumerate}
\item
Supersymmetry remains indeed unbroken up to micrometer scales,
but {\it only} in the bulk supergravity sector.
\item
On the brane, supersymmetry is broken at much larger scales
of order $TeV$. As in \cite{ant},
the brane is in fact assumed to be the
origin of supersymmetry breaking:
we start with a supersymmetric string compactification that is perturbed
by one or more  stable 3--brane solitons, parallel to the
4 uncompactified coordinates; those solitons are assumed to be non--BPS,
so they break all
supersymmetries (and not just half of them).
\item
In this situation, where supersymmetry
breaking originates in the matter sector (on the brane)
 and is transmitted to the gravity sector (the bulk), one naturally expects
the corresponding scales $m_{brane}$ and $m_{bulk}$ of supersymmetry breaking
to be related by
$m_{bulk}\sim (m_{brane}^2/ m_{Pl})$ \cite{ant}.
Thus, $m_{brane}$ is half--way between $m_{bulk}$ and $m_{Pl}$ on
a logarithmic scale (see figure 1). If $m_{brane}\sim TeV$, then
supersymmetry in the bulk supergravity sector remains
unbroken up to scales of order $m_{bulk}\sim meV$ (milli-$eV$).
\item
Supersymmetry breaking on the brane creates a huge brane vacuum energy
of order $(TeV)^{4}$. But as in the mechanism of Rubakov and Shaposhnikov \cite{rub},
it has been argued that this brane vacuum energy does not curve
the four--dimensional spacetime parallel to the brane and is instead
absorbed by the curvature transverse to the brane. 
\item
The only fields whose
vacuum energy
contributes to the  four--dimensional curvature 
(and therefore to the cosmological constant) are
then the bulk supergravity fields. Under two important
assumptions that will be stated, with bulk supersymmetry
being broken at $meV$ (milli--eV) scales this indeed gives
a Hubble constant 
of roughly the observed order of magnitude. 
\item
Given the values
of the Hubble constant and the Planck mass,
 we can reverse the argument and try to predict the precise masses
of syperpartners not only on the brane, but also in the bulk.
This mass spectrum of superpartners of the graviton in the milli--$eV$ range
 should be observable 
in measurements of Newton's constant in the micrometer range.
In our scenario,
these predictions depend on the detailed non-BPS soliton solution.
Therefore the spectrum
of supergravity masses should offer a window through which we
can probe what kinds of branes our string compactification contains.
\end{enumerate}
 
\begin{figure}[htb]
 \vspace{9pt}
\vskip5cm
 \epsffile[1 1 0 0]{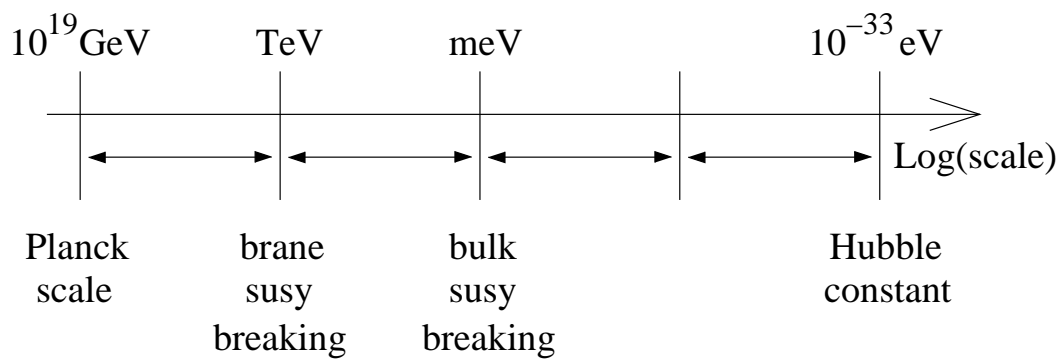}
\caption{Evenly spaced hierarchies on a logarithmic scale.}
\end{figure}

\noindent
In this note, these steps will be reviewed in more detail
and the following hidden assumptions and
objections against this scenario  will be
discussed:
\begin{enumerate}
\item
{\it The issue of boundary conditions:}
due to obstructions to finding a global solution for the
background geometry, it has been argued \cite{nil} that
the Rubakov-Shaposhnikov mechanism cannot be
extended to models with a single extra dimension \cite{kss}.
It will be argued that in order to avoid similar problems,
at least two extra dimensions are needed.
\item
{\it The issue of nearby curved solutions:}
it remains an open question whether
 supersymmetry in the bulk is good enough 
to ensure that the solution where {\it all} of the brane vacuum energy
is absorbed in the curvature {\it transverse} (rather than parallel)
to the brane is the stable one.
\item
It must be demonstrated in concrete examples
that $m_{bulk}$ is really in the milli--$eV$ range.
Even when this is the case, 
the bulk fields produce a curvature radius of the universe
that is too big (of solar system scales), unless
$\hbox{Str\ }M_{bulk}^2\ =\ 0\ .$
\item
It must be assumed that Kaluza Klein modes 
do not contribute to the
4--dimensional cosmological constant, e.g. because their mass splittings
might be very small.
\item
Our scenario can be ruled out experimentally if Newton's constant 
is observed to remain unchanged down to scales of
$10\ \mu m\ .$  This is more an advantage than a problem.
\end{enumerate}
Points 1 and 2 refer to the Rubakov--Shaposhnikov mechanism in
classical supergravity and are discussed in section 2.
Points 3--5 refer to one--loop supergravity corrections and
are discussed in section 3. Section 4 contains an (optimistic) conclusion.

That the Rubakov--Shaposhnikov mechanism should be 
combined with supersymmetry away from the brane to adress
the ``first cosmological constant problem'' (why $\lambda$ is zero \cite{wei2})
was -- as far as the author is informed -- first suggested in \cite{ver2}.
What was argued in \cite{schm} is that if this classical mechanism works, then 
one--loop supergravity corrections can also solve the ``second cosmological
constant problem'' (why $\lambda$ is tiny but nonzero), and that 
this could be tested by looking for 
superpartners of the graviton
 with Compton wavelengths in the micrometer range.

\subsection*{2. Classical Objections}\scs{2}

In this section we discuss the classical approximation to the 
bulk supergravity theory. 
The world--brane theory, though, is assumed to be treated to all orders
in and beyond perturbation theory.

Let us first make the setup more precise.
We assume that the $n$ extra dimensions are compactified on some
compact manifold ${\cal M}$. The brane itself is a soliton solution
of the supergravity equations of motion; its characteristic size
should be of order $\sqrt{\alpha'}$, the fundamental scale
in string theory. In this setup it is natural to identify 
$\sqrt{\alpha'}$ with the (inverse) scale $m_{brane}$ of
supersymmetry breaking in the Standard Model (on the brane),
since the brane is the origin of supersymmetry breaking.
There is one free parameter in our discussion,
$$\tau^2\ =\ {Vol ({\cal B})\over Vol({\cal M})}\ \sim\ 
{m_{brane}^2\over m_{Pl}^2}\ ,$$
the ratio of the volumes of ${\cal M}$ and of ${\cal B}$,
the ball in which the brane intersects ${\cal M}$. $\tau$ is
related to the hierarchy between the Planck scale and
the supersymmetry breaking scale in the Standard Model similarly as 
in \cite{wit2,dim}. Our scenario will predict 
$\tau\sim10^{-15}$. 
We do not attempt to explain the origin of this hierarchy;
at best, we hope to reduce the cosmological constant problem to
the hierarchy problem.

Non--BPS stable branes have been presented, e.g., in \cite{sen,por},
and in the present context in \cite{may}.
Although these branes are stable in the sense that they have no 
world--brane tachyons, they are not stable 
under variations of the moduli of the compactification
geometry. The models in \cite{iba,blu} might be viable,
provided the tachyons appearing there can indeed be interpreted as
Higgs fields.
Of course, if no truly stable non--BPS solitons can be found,
one can always make the ansatz that one of the 
BPS branes in a supersymmetric
string compactification that involves space--time filling
3--branes is perturbed out of the BPS ground state,
 then relaxes back slowly to the supersymmetric equilibrium.

Let us now discuss the mechanism that has been held responsible
for absorbing the brane vacuum energy in the bulk.
For the metric near the brane, we make the ansatz
$$ds^2\ =\ dr^2\ +\ e^{2\alpha(r)}\hat g_{\mu\nu}dx^\mu dx^\nu\ 
+\ \tilde g_{ab}dx^a dx^b\ ,$$
where $\mu,\nu\in\{0,1,2,3\}$, $a,b\in\{4,...,3+n\}$, $r$ is the distance
from the brane, $\hat g_{\mu\nu}$ is proportional to
the metric of the four--dimensional low--energy
effective theory and $\tilde g$ is the metric on
the compactification manifold ${\cal M}$.

The $4+n$--dimensional spacetime can be split into two regions:
the region near the brane, where supersymmetry is broken;
and the bulk region, where supersymmetry in the
four--dimensional sense (on slices parallel to but far away
from the brane) is assumed to be unbroken. The classical supergravity equations of motion
can be solved separately in both regions and can then be
matched at their boundary.

In the bulk region, unbroken four--dimensional supersymmetry implies
$$\hat R_{\mu\nu}\ = \ 0\ ,$$
at least in the absence of four--form field strengths or expectation
values of other supergravity fields.\footnote{If there was a 
four--form field strength $F^{(4)}$, one could have a supersymmetric solution
with $\hat R_{\mu\nu}\sim\ F^{(4)}_{\mu\alpha\beta\gamma}F^{(4)}_\nu{}^{\alpha\beta\gamma}$.}

As for the brane region, 
there should be a horizon near $r=0$, where
the classical supergravity approximation  breaks down.
We therefore restrict the discussion to a region $r>\epsilon$ 
where derivatives with respect 
to $r$ are small, with some
cutoff $\epsilon$. The issue of boundary conditions at $r=\epsilon$
will be discussed shortly.

To lowest order in derivatives with respect to $r$,
 the Einstein equations are
$$R_{mn}\ -\ {1\over2}g_{mn}R\ =\ \lambda(r)g_{mn}\ .$$
Here $m,n\in\{0,...,3+n\}$, and
$\lambda(r)$ is the cosmological constant in the vicinity
of the brane that is created by brane supersymmetry breaking
and by the flux that emanates from the brane in the case where the brane
carries some charge. 
We can split
\ba R_{\mu\nu} &=& \hat R_{\mu\nu}\ -\ e^{2\alpha}\hat g_{\mu\nu}
\ \{\Box^{(n)}\alpha\ + 4(\nabla^{(n)}\alpha)^2\}\\
R_{ab}&=&\tilde R_{ab}\ +\ 4\{\nabla_a\nabla_b\alpha
+\nabla_a\alpha\nabla_b\alpha\},
\ea
where $\Box^{(n)},\nabla^{(n)}$ act on the closed transverse space.
Plugging this into the Einstein equations
yields \cite{rub}:
\ba
\hat R_{\mu\nu}&=&k^2\hat g_{\mu\nu}\la{laura}\\
k^2&=& e^{2\alpha}\{-{2\over2+n}\lambda(r)
\ +\ 4(\nabla^{(n)}\alpha)^2\ +\ \Box^{(n)}\alpha\}\\
{\tilde R}_{ab}&=& -{2\over2+n}\lambda\tilde g_{ab}
\ -\ 4\{\nabla_a\nabla_b\alpha
+\nabla_a\alpha\nabla_b\alpha\}\ .\la{lisa}
\ea
$k^2$ is an integration constant which, by definition, is the
four--dimensional cosmological constant. If solutions to these
equations can be found for all $k$, then there is
a one--parameter family of possibilities, labelled by $k$, of what one can do
with
the brane cosmological constant $\lambda(r)$: at one end of
this one--parameter family, the transverse curvature described by
${\dot\alpha}$ and 
$\tilde R$ is zero and $\lambda(r)$ goes into the four--dimensional
cosmological constant $k^2$. At the other end, $k^2$ is zero
and the brane cosmological constant $\lambda(r)$ is completely
absorbed by the transverse curvature. 

This is the mechanism of Rubakov and Shaposhnikov \cite{rub}. They remarked
that they had merely geometrically reformulated the fine--tuning problem:
now the question is, why should $k$ be fine--tuned to be exactly
zero? But in our case, it has been argued that matching to the solution in the region  away
from the brane picks the solution with $k=0$ \cite{ver2}.

In the following we mention two potential
 problems with this step of the argument.
First, the issue of boundary conditions, and second,
the question whether supersymmetry in the bulk is
 good enough to select the solution
with $k=0$.

As for the first problem, if
we want to have a closed compactification manifold,
we must actually impose boundary conditions on $\alpha(r)$ and $\tilde g(r)$
both at $r=\epsilon$ and at large $r$. But there may be global
obstructions to our ability to impose such boundary conditions.

As an illustration of the type of obstructions that may occur,
consider the case of two extra dimensions, i.e., branes of
co--dimension 2. Those are analogous to cosmic strings, and they
typically produce a deficit angle $\gamma$ in the tranverse two--dimensional
metric $\tilde g$. The boundary condition at $r=\epsilon$ in this case
says that the deficit angle is
$$ \gamma\ \propto\ T\ ,$$
where $T$ is the brane tension. To get a closed geometry we
must have several branes with deficit angles $\gamma_i$.
The deficit angles must add up to $2\pi$ (or a multiple $n$ thereof),
so there is a sum rule, analogous to that in \cite{nil}:
$$2n\pi\ =\ \sum_i \gamma_i\ \propto\ \sum_i T_i\ .$$
It can be satisfied for BPS branes, where the deficit angle is
fixed by a BPS bound and often a rational fraction of $2\pi$.
But if we break supersymmetry on one of the branes, then the 
brane cosmological constant, and thereby the brane tension and the deficit
angle are out of control. Since only a discrete set
of brane cosmological constants can be absorbed in the transverse
curvature, there is no continuus ``Rubakov--Shaposhnikov modulus''
that can be adjusted to absorb this arbitrary brane tension.

However, there
does not seem to be any such global obstruction to absorbing
an arbitrary tension for branes of codimension 3 or more. In those cases,
the transverse geometry can simply react to an increase in
brane tensions by changing its shape and size.
The above discussion is over--simplified, because there is also 
the warp factor $\alpha(r)$ that enters the equations (\ref{laura}--\ref{lisa}). As demonstrated in \cite{rub} (there in the case of an $r$--independent $\lambda$),
if $\alpha$ is included, then there may be ranges of $\lambda$
where solutions with compact transverse manifolds exist even for $n=2$.

As discussed in \cite{nil}, however,
the situation is worse in the case of a single extra dimension.
In this case, either the fine--tuning problem is hidden
in the boundary conditions near a singularity,
or the extra dimension is noncompact.
We assume a compact manifold ${\cal M}$, so
let us assume in the following that the number of extra dimensions is
$$n\ \ge\ 2 \ $$
and that $k$ is indeed a modulus of the classical solution
that can be adjusted to zero.

But in this case we still have the second problem mentioned above.
There are nearby curved solutions with $k\neq0$.
What picks the solution with $k=0$? The argument that has been
given above for selecting $k=0$ is supersymmetry in the bulk.
But won't supersymmetry breaking on the brane backreact on the
bulk already at the classical supergravity level, thus invalidating this argument?

If there was no warp factor, this would certainly be the case.
The brane would want to have
$k$ of order $TeV$, while the bulk -- being supersymmetric --
would want to have $k=0$. The bulk -- being much larger --
dominates (equivalently, the $4d$ Newton constant is small), 
but the ``compromise'' between bulk and brane would still be $k\sim(mm)^{-1}$,
which would be way too big. Due to the warp factor, though,
 at least at the classical supergravity level the brane doesn't care,
while the bulk still wants to be supersymmetric.
So it seems the compromise could be $k=0$. 

But even if so, what about supergravity
loop corrections? The modulus $k$, like any modulus in a 
string compactification, will become an effective $4d$ field,
which  may receive a complicated potential from
loops of the bulk fields. Even if this potential is small
due to the smallness of Newton's constant, it
is not clear that the potential is such that
its minimum is at $k=0$.\footnote{Thanks to Tom Banks
for pointing this out; see also \cite{wit3}.} Unfortunately this remains
an open problem
of the  scenario outlined in section 1.

\subsection*{3. One--loop objections}

Having mentioned this problem, let us now assume for the sake of argument that
it can be resolved, and that
 the Rubakov--Shaposhnikov mechanism works in the
sense that all the brane vacuum energy gets absorbed in the bulk.
Let us then consider the one--loop vacuum energy in the
bulk sector and ask whether it really gives us
 the right order of magnitude for the cosmological constant.
In other words, assuming that the ``first cosmological constant problem''
is solved by the Rubakov--Shaposhnikov mechanism, do one--loop
supergravity corrections really
solve the ``second cosmological constant problem'' as suggested in \cite{schm}?

The one--loop vacuum energy in the bulk sector of the low--energy
effective theory vanishes as long as supersymmetry is unbroken in the bulk.
But supersymmetry breaking is transmitted from the brane
to the bulk. Intuitively, because 
of the small overlap of the supergravity wave 
functions with the brane,  the supersymmetry breaking scale in the 
bulk should be suppressed by the same volume factor as before:
\ba {m^2_{bulk}\over m^2_{brane}}\ \sim\ {\vol({\cal B})\over \vol({\cal M})}
\ \sim\ {m^2_{brane}\over m^2_{Planck}}\ .\la{maxi}\ea
To actually compute the bulk masses precisely, one should
find the eigenvalues of the various wave operators in the
gravitational background of the non--BPS soliton:
\ba
0&=&\Box\phi\ =\ \Box^{(4)}\phi\ +\ \tilde g^{ab}\nabla_a\nabla_b\phi\la{rosa}\\
0&=&\slash\hskip-2.5mm D \psi\ =\ \slash\hskip-2.5mm D^{(4)}\psi\ +\ \slash\hskip-2.5mm D^{(n)}\psi
\la{rosi}\ea
and the same for the Rarita--Schwinger operator,
etc. The lowest eigenvalues of the wave operators on ${\cal M}$
are then the masses $m_0, m_{1/2}, m_{3/2}, ...$ of the superpartners
of the graviton.

At least in simple toy models, such masses are indeed
typically suppressed by the volume factor. E.g., consider the case of a 
scalar field in one extra dimension (without gravity), with a Schr\"odinger--type
equation of motion
$$\phi''\ +\ V(x)\phi\ +\ m^2\phi\ =\ 0\ .$$
If we make the toy model ansatz
$$x\ \in\ [0,L]\ ,\ V(x)\ = \ \Theta({\epsilon\over2}-|x-{L\over2}|)\ ,$$
then the zero mode will be of the form 
\ba\phi(x)&\sim& \cos\ mx \ \ \ \hbox{for}\ \ \ x<{L\over2}-{\epsilon\over2}\
\ \ \hbox{with}\ \ \ m^2\ =\ {\epsilon\over L}\\
&\sim& \cos\ m(L-x)  \ \ \ \hbox{for}\ \ \ x>{L\over2}+{\epsilon\over2}.\ea
The term $(\epsilon/L)$ is indeed the volume factor. Similar toy models
in higher dimensions yield similar results. Generalizing 
the discussion to the three--brane geometry is left for the future;
 let us just note that it seems important to ``get out of the throat'',
i.e. to not only restrict the discussion to the near--horizon limit
$AdS_5\times S^5$.

In the introduction, the contribution to the cosmological constant
from a single massless bosonic degree of freedom has been given,
 assuming a short--distance
cutoff $\Lambda$. In the presence of masses $m_i$, this formula generalizes 
(expanding in powers of ${m_i\over\Lambda}$) to
\ba\lambda\ =\ 3\Omega_\Lambda\ H^2\ =\ {1\over 2\pi m_{Pl}^2}\sum_i(-1)^{F_{i}}
\{\Lambda^4\ +\ m_i^2\Lambda^2\ -\ m_i^4\log{\Lambda^2\over m_i^2}\ +\ ...\}\ ,
\la{mary}\ea
where $F_i$ is the fermion number. In our case, $m_i$ are the masses of the supergravity
degrees of freedom labelled by $i$ (i.e., $i$ runs from 1 to 256 in the case of maximal supergravity).

The $\Lambda^4$ term vanishes, since in a theory with spontaneously
broken supersymmetry such as this one (where the classical soliton solution
does not respect the supersymmetry of the underlying Lagrangean)
there is an equal number of bulk bosons and bulk fermions. What about the 
$\Lambda^2$ term? It is proportional to the supertrace of the
square of the mass matrix in the
bulk supergravity sector. The question is then, is
$$Str\ {\cal M}_{bulk}^2\ =\ 0\ ?$$
We will come back to this shortly.
For the moment, let us assume that
we can find a situation where this is the case.
Then the relevant term for the cosmological constant is the last one in
(\ref{mary}).
Up the the logarithmic factor, which is of order 100 if we identify the
cutoff $\Lambda$ with the fundamental scale $m_{brane}$ of the
theory, this last term indeed reproduces the four equal steps on the logarithmic
scale of hierarchies drawn in figure 1, using (\ref{maxi}).

In the absence of concrete classical solutions for
non--BPS stable branes, we can not yet work out precise supergravity 
masses from (\ref{rosa},\ref{rosi}) etc. However,
working out the numbers for some randomly picked
patterns of supergravity masses, the majority of predictions are
 in the range \cite{schm}
$$m_{brane}\ \sim\ 2-6\ TeV\ \ ,\ \ \ \ m_{bulk}\ \sim\ .5-5\ meV\ \ .$$
Both of these ranges are the subject of planned measurements.
The latter prediction is not yet ruled out by experiment \cite{zwi}.
Since the contributions to $\lambda$ go like the 4th power of the mass,
it is quite possible that one superpartner of the graviton
 (whichever is the heaviest one)
 produces the lion's share of the cosmological constant.
If so, then this heaviest superpartner should not be heavier than 
$5\ meV\ \sim\ (40\ \mu m)^{-1}$, otherwise it
would produce a Hubble expansion that is too fast.

If $Str\ {\cal M}_{bulk}^2$
was nonzero, the curvature radius of the universe would be
at the third mark in the diagram in figure 1. 
This roughly corresponds to the distance between the planets
in the solar system, which is obviously too small.
In theories where global supersymmetry is spontaneously broken,
$Str\ {\cal M}_{bulk}^2$ is known to be zero at tree level.
But here we have {\it local} supersymmetry and there is no guarantee
that the supertrace vanishes. So $Str\ {\cal M}_{bulk}^2=0$
is an important consistency criterion that must be checked for each
non--BPS soliton background individually. However, since finding 
backgrounds with $Str\ {\cal M}_{bulk}^2=0$ involves no fine--tuning,
this may in fact be a blessing: it might allow one to throw out most
of the candidate non--BPS solitons.
Situations with $Str\ {\cal M}_{bulk}^2=0$ have been considered in \cite{fer}.

Even when  $Str\ {\cal M}_{bulk}^2\ =\ 0$ at tree level,
it will generally not be zero after loop corrections;
however, since we are concerned  only
about supergravity loops, these loop corrections are 
 suppressed by additional powers of Newton's constant and should therefore
be small.

What {\it is} a  potentially fatal
problem with our scenario, though, is the following. So far, we 
have only considered the contribution
of the Kaluza--Klein zero modes to both $Str\ {\cal M}_{bulk}^4$
and $Str\ {\cal M}_{bulk}^2$. But 
the higher Kaluza--Klein modes should also contribute.
We must assume that $Str\ {\cal M}_{bulk}^2\ =\ 0$ also
for the Kaluza--Klein modes, and moreover, that they do not contribute to
 $Str\ {\cal M}_{bulk}^4$ -- either because their
mass splittings are tiny, or because of some other unexpected feature of the
effective four--dimensional Kaluza--Klein gauge theory.\footnote{E.g.,
one might speculate that the Kaluza--Klein modes are confined,
the confined charge being Kaluza--Klein momentum; Confinement of
Kaluza--Klein momentum has been discussed in \cite{gro}, and
in a three--dimensional context in \cite{schm4}.}

Boldly assuming that this is the case, here are the steps
that would have to be followed in order to make the above predictions
for future experiments more precise.

\begin{enumerate}
\item
Make a list of string compactifications involving
 non--BPS stable 3--branes with co--dimension at
least two, and find their
supergravity solutions.
\item
For each such compactification, compute the supergravity masses
as a function of the volume factor $\tau^2$,
by finding the eigenvalues of wave operators on ${\cal M}$
in these backgrounds.
This will probably only be possible in an $\alpha'$ expansion.
\item
Given these eigenvalues, compute $Str\ {\cal M}_{bulk}^2$ 
for each compactification. Compactifications with $Str\ {\cal M}_{bulk}^2\neq0$
must be thrown out. Hopefully this leaves only a tiny set of
canidate compactifications.
\item
Compute $Str \ {\cal M}_{bulk}^4(\tau)$ and 
then use the observed ratio between the 
Planck and the Hubble constants to
calibrate $\tau$.
\item
From the supergravity masses, compute a ``running Newton constant''
$G_N(\mu)$: the effective Newton constant at micrometer scales
should change due to the contribution from the exchange of
particles such as the dilaton. Then compare the curve $G_N(\mu)$
with short--distance measurements
of gravity in the micrometer range. 
\end{enumerate}
If no deviations from
Newton's constant are found at all down to scales of, say, $10\mu m$,
it is probably safe to say that our explanation for the small
cosmological constant is wrong.
On the other hand, if such
deviations {\it are} found, they could be the beginning
of an exciting period  of experiment--driven superstring phenomenology.
The curve $G_N(\mu)$ would encode information about
the supergravity masses, and thereby about
the type of non--BPS solitons that the ``real--world''
 superstring compactification contains.

\vskip3cm
\subsection*{4. Conclusion}\scs{4}

At least if the recent measurements of the cosmological constant
can be trusted, then the size of the cosmological constant is such
that it could be produced by a supermultiplet of particles with
supersymmetry being broken in the milli--$eV$ range.
Whatever these particles are, they must be fundamentally different
from the Standard Model particles, because they contribute to the
cosmological constant, while the Standard Model particles obviously
don't.

The non--BPS brane world scenario provides precisely such a fundamental
difference between the Standard Model particles, which live
on the brane, and other particles (the supergravity particles),
which live in the bulk. In addition, it provides a mechanism
(the Rubakov--Shaposhnikov mechanism) which could at least in
principle do the job of soaking up the Standard Model vacuum energy.

Most strikingly, the supersymmetry breaking scale of order milli--$eV$
comes out naturally, if one assumes Standard Model supersymmetry
breaking in the $TeV$ range. 
This feels a bit like pieces of a puzzle falling into place.
It may be part of the solution of the cosmological
constant puzzle. 

But obviously, what has been discussed here cannot be the whole
story because of the problems mentioned --
most notably, reasons are missing why the potential
for the modulus $k$ should have its minimum at zero, and
why higher Kaluza--Klein modes should be ignored.

Perhaps the best feature of our scenario is that it
predicts contact with experiment.
Clues about whether and how these problems can be solved will thus hopefully come
from measurements of Newton's constant in the range between
$10$ and 1000 $\mu m$, such as \cite{exp}.

\vskip2cm
\noindent
{\bf Acknowledgements:}

\noindent
I would like to thank T. Banks, S. Kachru, H.P. Nilles, E. Witten
and especially P. Mayr for helpful discussions and/or criticism. 
I also thank members of
the groups in Halle, Berlin, Bonn, Warsaw and at Cern
for comments.

\end{document}